\documentstyle[aps,twocolumn]{revtex}

\begin{document}

\draft


\title{
Comment on: 
 ``Trace anomaly of dilaton coupled scalars in two 
dimensions'' }


\author{ W.~ Kummer\thanks{e-mail: wkummer@tph.tuwien.ac.at}, 
H. Liebl\thanks{e-mail: liebl@tph16.tuwien.ac.at}
       and D.V.\ Vassilevich\thanks{Permanent address:
Department of Theoretical Physics,
   St.\ Petersburg University, 198904 St.\ Petersburg,  Russia.
e.mail: vasilev@snoopy.phys.spbu.ru}}


\address{Institut f\"ur
    Theoretische Physik, Technische Universit\"at Wien, Wiedner
    Hauptstr.  8--10, A-1040 Wien, Austria}

\maketitle

\begin{abstract}
The trace anomaly for nonminimally coupled scalars in spherically 
reduced gravity obtained by Bousso and Hawking \cite{BH1} is
incorrect. We explain the 
reasons for the deviations from our correct (published) result which 
is supported by several other recent papers. 
\end{abstract}

The computation of the flux of Hawking radiation from the trace 
anomaly has a long history \cite{chr77}. For the 
case of gravity, spherically reduced to 1+1 dimensions, somewhat 
surprisingly, up to very recent times only calculations existed which 
were based on minimal coupling to (massless) scalar fields. However, 
spherical reduction for the matter interaction leads to an additional 
factor related to the dilaton field. Besides, also the integration 
measure for the scalars in the path integral is to be modified by the 
same factor.

This problem is closely related to our recent work on minimally 
coupled scalars in 2d Hawking radiation \cite{LVA} for a 
very general class of 2d models with one horizon and various global 
properties of the geometric background     \cite{KKL}. 
Therefore in \cite{KLV-mod-phys} we treated the 
most general case of the gravity models of ref.\cite{KKL}, even 
allowing for a measure for a scalar field with arbitrary dependence on 
the dilaton field. As pointed out by us already in \cite{KLV-mod-phys}, 
the 
trace anomaly for the special case of spherically reduced gravity 
differs from the --- at that time unpublished ---  result of Bousso and 
Hawking \cite{BH1}. \\

First of all the authors of \cite{BH1} use the 2D path integral
measure for scalar fields rather than the spherically reduced 
4D diffeomorphism invariant measure. Let the reduced action read

\begin{equation}
S=\int d^2x \sqrt{-g}fAf
\end{equation}
where $g$ is determined from the 2D metric obtained by spherical
symmetric reduction of the 4D line element

\begin{equation}
ds^2=g^{\mu \nu}dx_{\mu}dx_{\nu}+e^{-2\phi}d\Omega^2 \quad.
\end{equation}
The normalization condition for eigenmodes is 
\begin{equation}
\langle f_{\lambda} f_{\lambda'} \rangle = 
\int d^4x \sqrt{-g}e^{-2\phi}f_{\lambda} f_{\lambda'} =
\delta_{\lambda,\lambda'} \quad .
\end{equation}
Substituting the $s$-wave modes 
$(f_{\lambda} f_{\lambda'})$ one obtains

\begin{equation}
\langle f_{\lambda} f_{\lambda'} \rangle \propto 
\int d^2x \sqrt{-g}e^{-2\phi}f_{\lambda} f_{\lambda'} =
\delta_{\lambda,\lambda'} \quad .
\end{equation}
Therefore in the standard $\zeta$-function or heat kernel methods one
must first introduce new fields $\tilde{f}=e^{-\phi}f$ obeying
the familiar 2D normalization condition
\begin{equation}
\langle \tilde{f}_{\lambda} \tilde{f}_{\lambda'} \rangle =
\int d^2x \sqrt{-g}\tilde{f}_{\lambda} \tilde{f}_{\lambda'} = 
\delta_{\lambda,\lambda'} \quad .
\end{equation}

This is equivalent to using the $\zeta$-function of 
$\tilde{A}=e^{\phi}Ae^{-\phi}$ in the definition of the effective
action and stress energy tensor. This correct path integral
measure was used in \cite{mwz} thus leading to results coinciding
with ours. Bousso and Hawking use $\zeta_{A}$ instead of
$\zeta_{\tilde{A}}$.
Hence their results are not related to spherically reduced
4D quantum matter.

Also the method used by the authors of \cite{BH1} to calculate
the anomaly for a given operator is ambiguous.
The simplest way to define the trace of the stress energy tensor
is to use behavior of the effective action $W$ under 
infinitesimal transformations 
$\delta g_{\mu\nu} = \delta k(x) g_{\mu\nu}$
leading to 
\begin{equation}
\delta W=\frac 12 \int d^2x \sqrt g\delta g^{\mu\nu}T_{\mu\nu}
=-\frac 12 \int d^2x \sqrt g \delta k(x)T_\mu^\mu (x) \quad .
\label{T}
\end{equation}
However, if one applies only a {\it global} scale transformation instead
-- as Bousso
and Hawking do -- one can only obtain an integral of $T_{\mu}^{\mu}$.
This obviously is not enough to fix $T_{\mu}^{\mu}$.
Bousso and Hawking correctly admit an ambiguity in their approach
which, however goes even beyond total derivatives.
Here it must be stressed that such an ambiguity is only a weakness of their
technique \cite{BH1} -- it does {\it not} reflect any physical
ambiguity.
As soon as the action and the path integral measure are fixed 
$T_{\mu}^{\mu}$ is determined uniquely by a {\it local} 
$\delta k(x)$
at least if the same
(zeta -) regularization is employed.
The correct result for spherically reduced gravity is
\begin{equation}
T_\mu^\mu =\frac 1{24\pi} (R-6(\nabla \phi )^2 +6\Box \phi ) 
\label{T3}
\end{equation}
whereas in \cite{BH1} the last term in (\ref{T3})
reads $-2\Box \phi$.
In the published version of \cite{BH1} we also now noticed a reference to 
the work of Chiba and Siino \cite{chiba} which we had not been aware of 
before. Also that reference already contains the correct trace 
anomaly just as the even earlier work of Mukhanov et al. \cite{mwz}.
Searching now the literature on this subject for the period 
after \cite{BH1} had been put into the electronic archives, 
we realized the existence of 
several new papers \cite{ichi,chiba} which all support our calculation. 
Therefore it is difficult to understand why the mistake in \cite{BH1} had 
not been corrected by the authors themselves, and why the same 
incorrect expression even had been used again in a later paper \cite{BH2} 
of the same authors. \\

We regret very much that several attempts to settle this issue by 
correspondence did not lead to a correction of \cite{BH1} in the final 
printed version. These authors also still do not refer to our paper 
\cite{KLV-mod-phys}
although it had even appeared in print in the 
meantime. Equally regrettable is the consequence that recent 
work by other authors almost exclusively refers to the incorrect 
result of \cite{BH1} and not to the correct one of 
\cite{KLV-mod-phys,ichi,mwz,chiba,noj97a}
among which our treatment is the most general one. 

\vspace{0.5cm}

{\large\bf Acknowledgement}

We thank Andrei Zelnikov for drawing our attention to ref. \cite{mwz}.
This work has been supported by Fonds zur F\"orderung der
wissenschaftlichen For\-schung (FWF) Project No.\ P 10221--PHY.  One
of the authors (D.V.) thanks the Russian Foundation for 
Fundamental Research, grant
97-01-01186, for financial support.

\end{document}